# Impact of humanity on climate change

## Vladimir Kh. Dobruskin

Beer Yacov, Israel; Email: dobruskn@ gmail.com

The thermodynamic approach shows that the total energy produced by humanity disrupts the thermal balance of the planet and causes counteraction, that is, climate change, which can slow down the progress of civilization. This outcome not only confirms the consensus on the role of humanity in climate change, but also quantifies it. The calculation of the average temperature of the Earth is based on its energy balance and includes the application of the Stefan-Boltzmann equation. The confusion in the use of the equation was discovered and clarified; the boundaries of application of the Law was outlined. The application of the Stefan-Boltzmann equation shows that the energy produced by humanity is responsible for about half of the magnitude of global warming. This result should have a strong impact on environmental decision-making.

**1**. **Aim of the research**

Thousands of researchers around the world are involved in finding solutions to problems related to climate change and global warming. Respected scientists warn humanity about the approaching threat and call for urgent measures to prevent dangers [1-5]. There is a consensus that it is human activity that destroys the environment. The most common view is that global warming is caused by the following chain of events: the burning of organic fuel increases the concentration of $CO_2$ in the atmosphere, which, due to the greenhouse effect, leads to an increase in temperature of the planet.

The contribution of the greenhouse effect to the climate of the planets of the Solar system is undeniable and has been proven by astrophysicists. Experimental data on the increase in the concentration of $CO_2$ and on the increase in Earth's temperature over the past 150-200 years are also indisputable. The graphs of these changes versus time look very similar. However, all this cannot be considered as proof of one-to-one correspondence between the phenomena. There is no evidence that the greenhouse effect is the sole cause of global warming [6].

The purpose of this study is to show that the energy produced by mankind is partly responsible of global changing. An integral part of this work is the use of the Stefan-Boltzmann law to estimate the radiation of Earth's surface. An analysis of the practice of using the law has shown that there is confusion related to the concept of the average temperature of the Earth. For this reason, this study consists of two parts: first, section 2 explains the essence of confusion and outlines the boundaries of the correct application of the Law, and then the impact of humanity on global warming is examined (section 3).

## 2. Confusion in the application of the Stefan-Boltzmann law

The Stefan-Boltzmann law relates the temperature of the blackbody, $T$, to the amount of power, $P$, it emits per unit area:

$$P = \sigma T^4, \tag{1}$$

where $\sigma = 5.6704 \times 10^{-8}$ W/m² °K⁴ is the constant [7]. This law, like the underlying Planck's law, is valid for a black body with uniform $T$, which is also in thermal equilibrium with the environment. When applying the equation (1), the Earth is approximately considered as a black body, but the fact that it is not in thermal equilibrium either by itself or with the environment is ignored. The latter leads to the serious error in using equation (1), the root of which lies in the confusion in concept of "the average temperature of Earth".

The details of calculating the average temperature of the Earth can be found in the specialized literature [6]. It is important that everything boils down to calculating the arithmetic mean using the adopted grid of measuring stations on the earth's surface:

$$T_a = \sum_1^n T_i / n, \tag{2}$$

where $n$ is the quantity of grid nodes and $T_i$ is the temperature of $i$-th node. The subscripts "$a$" (arithmetic) and "$p$" (power) will be used to distinguish values corresponding to either arithmetic or power-law means of temperature, respectively. There is no device for measuring the average temperature of Earth; the latter is simply a convenient statistical indicator that proves very useful for assessing climate change on a global scale [6].

However, the use of $T_a$ in equation (1):

$$P_a = \sigma (\sum_1^n T_i / n)^4, \tag{3}$$

leads to an incorrect estimation of the Earth's radiation, since the Earth is not an equilibrium body with uniform temperature. This inaccuracy can be eliminated by applying equation (1) to local regions with constant temperature and subsequent summation across the planet. Then the power radiated by the square meter of planet`s surface is given by:

$$P_p = \sigma\left(\sum_1^n T_i^4/n\right), \tag{4}$$

where

$$T_p = \sqrt[4]{\sum_1^n T_i^4/n} \tag{5}$$

is the correct value of the average temperature. Note that $T_p$ is the mean of fourth powers, whereas $T_a$ is the arithmetic mean (mean of first powers). The local temperature varies over the surface of the earth from the coldest temperature, -89.2°C, in Vostok, Antarctica to the hottest value ever recorded on Earth, 70.7°C in the desert of Iran [8]. For these two regions, $T_p$=294.9°K, whereas $T_a$ = 263.9°K. Of cause, it is the extreme case, and for the Earth as a whole the difference between $T_p$ and $T_a$ will be smaller, but it never disappears at all.

The distinction between $T_a$ and $T_b$ follows from a mathematical inequality [9, 10]

$$\left(\frac{x_1+x_2+\cdots+x_n}{n}\right)^\alpha \le \frac{x_1^\alpha+x_2^\alpha+\cdots+x_n^\alpha}{n}, \tag{6}$$

where $x_1, x_2, \ldots, x_n > 0$ and $\alpha > 1$. If we replace $x_i$ by $T_i$, set $\alpha = 4$, and extract the root of the fourth degree, we get

$$T_p \ge T_a \text{ and } P_p \ge P_a. \tag{7}$$

The equality could occur only if all regions have the same temperature, that is, the planet itself would be in equilibrium. The latter is impossible in the Solar system, however equation (1) with $T=T_a$ corresponds to just such an unrealistic case. An analogy can be drawn: the distinction between $T_a$ and $T_p$ resembles that between the average velocity and the root mean square velocity of molecules in kinetic theory of gases [11].

As a result, we come to the conclusion that (a) using the arithmetic mean temperature $T_a$ in equation (1) is conceptually incorrect and (b) the reverse action - calculating $T$ for a given $P$ - gives the value $T_p$ (and not $T_a$!). This conclusion should be taken into account when analyzing the role of various factors in global warming. The following is the example of the confusion.

The planet surface receives from Sun $P_s$=239 $W/m^2$ [12]. According to equations (4) and (5) one square meter of a black body radiates this amount of energy at $T_p$=255 °K, whereas the average temperature of earth is $T_a$=288 °K. The difference of temperatures, 33°K, is attributed to the greenhouse effect [13]. Strictly speaking, there is the error here, since 33°K is difference between $T_a$ and $T_p$, which differ from each other by definition (compare equations (2) and (5)). It is inaccurate to calculate the difference between $T_a$ and $T_p$, and the correct conclusion can be made only by comparing values marked with the same indices. In the case of temperature changes - either $\Delta T_a$ or $\Delta T_p$, and not a mixture of values with different indices.

## 3. Impact of human activity on global warming

### 3.1 Progress of civilization in the language of physics. Ideal and real states of Earth

The average temperature of Earth is determined by thermal balance, that is, how much energy the planet receives and how much it radiates back into space over the same period of time. Details can be found in the specialized literature [12]. It is worth emphasizing that the thermal equilibrium is usually considered for the unhabituated planet. To quantify the impact of humanity on climate change, a thermodynamic approach was proposed [14] based on the Kardashev idea. Considering hypothetical celestial civilizations, Kardashev suggested to evaluate development of the civilization by the amount of energy, $E$, it is able to use [15]. The idea of Kardashev can be basis for translating the concept of "human activity" into the language of physics: no human activity is possible without the use of energy. The amount of energy produced by humanity is in unambiguous accordance with the level of civilization. The more energy is generated, the higher the level of development of civilization and the stronger the impact of humanity on the environment. Nevertheless, the application of Kardashev's idea needs to be clarified.

For a long period of our history, until about 1800 year, the influence of civilization on the climate was insignificant, and humanity existed in harmony with nature - with clean air and drinking water, green forests, clean rivers and oceans, and so on. This state of the planet will be called the ideal state; it practically does not differ from the state of the uninhabited Earth. The average annual temperature has maintained at about 15°C. Obviously, everyone would prefer to live in such an environment. *It can be argued that in the absence of a developed civilization, such a state would remain virtually unchanged up to the present days and even in the near future.*

Unlike an ideal planet, anthropogenic evolution continues and accelerates on Earth, civilization continuously produces energy, and this energy inevitably disrupts the thermal balance that existed on the planet in its ideal state. Due to the thermal inertia of the oceans and the slow reaction of other elements of the climate system, it takes centuries for the climate to reach an equilibrium state [16]. For this reason, the real planet is not in equilibrium: its current state can only be considered as an indicator of what the displacement of the equilibrium of an ideal planet leads to.

Let's compare these two states: 1) the ideal state of the Earth and 2) the real state of the planet with an evolving civilization. According to the Le Chatelier principle [17], the dynamic equilibrium in the system is maintained until the conditions to which it corresponds are violated. If the latter occurs, the equilibrium position shifts to counteract the change; the process continues until a new equilibrium is reached. This is exactly what happens with the ideal system, which begins to turn into a real system. The reaction of the planet is really aimed at countering the destructive effects of the energy of civilization; it causes such changes in the environment (climate) that suppresses human activity and could slow down the progress of civilization in the future. Hence, the thermodynamical approach shows that the energy of civilization disrupts the thermal balance of the planet and causes opposition to human activity, which results in a climate change. This outcome not only confirms the consensus on the role of humanity in climate change, but also quantifies it.

### 3.2 Energy of civilization

The second law of thermodynamics states that the efficiency of any real process is less than 100%: part of the energy is dissipated in the environment in the form of heat, the rest, free energy, can be used for humanity needs. According to the first law, the free energy also cannot disappear, and in the process of producing useful work, it is spent on overcoming resistance and also turns into heat. Thus, on the scale of the Earth, all the energy produced by civilization, hereinafter referred to as civilization energy (EC), is inevitably converted into heat; it is transferred to environmental molecules (mainly atmospheric molecules), increasing their kinetic energy and temperature. Air molecules (N$_2$, O$_2$ and Ar) remain on Earth due to gravity, and the EC is accumulated on the planet. It can be said that the scattered EC eventually turns into the heat of civilization. However, we will not introduce a new term, since it is clear from the context what is being discussed. Therefore, the total thermal energy of civilization, $E_t$, for the time interval is summed up:

$$E_t = \sum_{i_1}^{i_2} E_i, \qquad (8)$$

where $E_i(i)$ is the annual energy production; $i_1$ and $i_2$ are the beginning and end of the interval, respectively.

For an uninhabited planet, the incoming energy of solar radiation is equal to the outgoing energy of terrestrial radiation. It follows from the constancy of the average temperature over the historical period. We can assume that for the inhabited earth, this ratio also remains almost constant for the time interval under discussion. On the contrary, the EC is summed up, and its influence on temperature is constantly increasing. Hence, the EC is the additional energy that the planet receives from the developing civilization.

This key point needs to be clarified. The total energy production in 2018 year was 171,240 TWh [18], and hence about $6.16 \times 10^{21}$ *J* for current *decade* (2013 ÷ 2022). Taking into account the Earth surface area, $5.10 \times 10^{14}$ m$^2$, for this decade human activity adds in average 0.38 *W*/m$^2$ (see eq.8). The sources of this energy - fossil fuels, radioactive substances, winds, rivers and tides - exist in the nature irrespective of either the earth is habituated or no. Why does the energy obtained from these resources affect the thermal balance of a planet with a developed civilization? and why are these resources neglected in the case of an uninhabited planet? *The underlying reason is a speed of energy liberation.* In nature, these sources slowly disperse their energy, while humanity forces them to liberate free energy within seconds. The ratio of speeds may reach the order of millions and much more. Below there are two illustrations. In nature, the energy of rivers is slowly dispersed over thousands of kilometers of riverbeds, whereas on hydroelectric dams it is released within a few seconds. Fossil fuels and radioactive substances have been around for millions and billions of years, but they are rapidly releasing energy in power plants. Obviously, if the speed of natural processes is million times slower, the contribution these sources to the heat balance of the unhabituated planet is neglectable – the order of $0.38 \times 10^{-6}$ W/m².

When considering the impact of humanity on climate change, the total EC determines environmental changes, while the progress of civilization is determined only by the free energy of civilization. As an illustration, if someone burns one ton of coal on a bonfire, the damage to the environment will be the same as if it was used in a coal-fired power plant. But in the second case, in the course of the process, free energy can be used for the development of civilization.

### 3.3 Temperature rise

The question arises: what the growth of temperature, $\Delta T_c$, is caused by the energy of civilization? The answer to this question is given by the Stefan-Boltzmann equation. For its application, it is necessary to convert the annual energy production $E(i)$ into the average annual power of civilization per m², $\varepsilon(i)$. Obviously, $\varepsilon(i)=E(i)/kS$, where $k=5.616 \times 10^6$ is the number of seconds per year and $S=5.1 \times 10^{14}$ m² is the Earth's surface area. Then, the temperature growth in $i$-th year is equal to

$$\Delta T(i) = \sqrt[4]{\frac{P_s + a + \varepsilon(i)}{\sigma}} - \sqrt[4]{\frac{P_s + a}{\sigma}}, \qquad (9)$$

$a$ is the EC contribution to $P_s$ up to $i$-th year. The total growth of temperature, $\Delta T_t$, for time interval from $i_1$ to $i_2$, then is equal to

$$\Delta T_t = \sum_{i_1}^{i_2} \Delta T(i) \qquad (10)$$

Since $a < P_s$, $a$ in eq. (9) may be ignored. Substitution eq. (9) in eq. (10) after some algebra leads to

$$\Delta T_t = \sqrt[4]{\frac{P_s + \sum \varepsilon(i)}{\sigma}} - \sqrt[4]{\frac{P_s}{\sigma}}, \qquad (11)$$

where $\sum \varepsilon(i)$ is the sum of $\varepsilon_i$ for the time interval under discussion. Since $\sum \varepsilon(i) \ll P_s$, the Taylor expansion of equation (9) leads to the following approximation:

$$\Delta T_t \approx 16.2008 \times [\sum \varepsilon(i)]/P_s^{0.75}, \qquad (12)$$

where 16.20008 °K/W$^{0.25}$ is the coefficient. Equation (12) shows that the growth of $\Delta T_t$ is proportional to the energy of civilization $\sum \varepsilon(i)$. Experimental data and calculated values related to the several time intervals are presented in Table 1.

Table 1. Impact of humanity on global warming

| Time interval, years | Energy of civilization, [18] | | Global warming, $\Delta T$,°C, | Contribution of EC to $\Delta T$, eq.(11) | |
|---|---|---|---|---|---|
| | $E_t$, (eq.8), TWh | $\sum \varepsilon(i)$, W/m$^2$ | | $\Delta T_t$ °C | % |
| 1985-1994 | 1.04×10$^6$ | 0.23 | 0.12[20] | 0.06 | 50% |
| 2013-2022 | 1.71×10$^6$ | 0.38 | 0.20 [19] | 0.10 | 50% |
| 1950-2020 | 6.25×10$^6$ | 1.40 | 0.82[20] | 0.37 | 45% |

Recall that $\Delta T_t$ is the growth of temperature only due to the energy produced by humanity, that is without taking into account other reasons such as, for example, an increase of concentration of $CO_2$ and greenhouse effect. In fact, the calculation answers the question: how much should the temperature of the planet rise in order to throw an additional EC in the form of radiation into space and achieve a new equilibrium. It is assumed that all other factors affecting temperature remain unchanged. Comparing $\Delta T_t$ with the experimental values of global warming $\Delta T$, one can conclude that about half of global warming is caused by the energy of civilization.

### 3. 4 The difference between the proposed approach and the traditional one

When there is a thermal balance of the earth's surface,

*energy received by the surface = energy radiated from the surface into space,*

the surface temperature remains constant. The term "surface" refers to the radiation surface in the Stefan-Boltzmann law, which is close to the geometric surface of the planet where human activity is concentrated. Obviously, the amounts of energy on the both sides of the balance refer to the same time interval, usually it is energy per second (power). If the left side of the balance begins to exceed the right, the temperature increases and vice versa, while the new balance is established. The difference between the proposed approach to global warming and the traditional one becomes obvious if we consider the thermal balances of these models.

The supposed approach proceeds from assumption that the energy of civilization is the additional energy to the solar radiation, which increases the left side of the balance. It does not matter for this model whether the energy source is renewable or non-renewable: one joule received from a wind farm has the same effect as one joule from a coal-fired power plant. We do not address the propagation of rays in the atmosphere, the absorption of infrared radiation by greenhouse gases and back emission to the Earth's surface (greenhouse effect) and proceed from the approximation that the contribution of these phenomena, the right part, remains unchanged.

The traditional approach, on the contrary, considers the left side unchanged and believes that the energy radiated into space decreases due to the growth in the concentration of carbon dioxide from 280 ppm to 400 ppm and a subsequent increase in back radiation of exited molecules of $CO_2$. On the Earth's scale, the greenhouse effect has been going on for billions of years, while the current global warming has been going on for only a few decades. It is possible that the contribution of the greenhouse effect in recent decades has been somewhat overestimated, and then both approaches are correct.

4. **Conclusion**

There are three implications from this study that require attention:

(1) There is an important practical difference between theories regarding the mitigation of climate change: the new approach recommends a reduction in global energy production, while the traditional approach focuses on reducing emission of carbon dioxide.

(2) Due to the generality of thermodynamics, it is impossible to neglect or circumvent its basic concepts. Albert Einstein wrote [21] about the classical thermodynamics: "It is the only physical theory of universal content that, I am convinced, will never be overthrown, within the framework of applicability of its basic concepts." Indeed, it is impossible to preserve the environment (nature and climate) as they were 200-500 years ago, and at the same time preserve the system (civilization) in its current state. If both approaches are correct, then it would be advisable, along with the measures already taken, to focus on reducing the energy consumption. This conclusion can have a strong impact on environmental decision-making.

10(3) The intermediate result of the study concerning the use of the Stefan-Boltzmann law is also noteworthy because it eliminates confusion that apparently went unnoticed during more than a century of application of the law.

**References**

1. Meadows, D.H., Meadows, D.L., Randers, J.W., and Behrens III, W.W. (1972). The Limits to Growth. New York: Universe Books.

2. Meadows, D.H., Randers, J. and Meadows, D.L. (2004). The Limits to Growth: The 30-Year Update. Vermont: Chelsea Green Publishing Co.

3. Scientist Statement: World Scientists' Warning to Humanity. (1992). Union of Concerned Scientists (ucsusa.org); W. J. Ripple, C. Wolf, T. M. Newsome, M. Galetti, M. Alamgir, E. Crist, M. I. Mahmoud, W. F. Laurance, 15,364 scientist signatories from 184 countries.

4. World Scientists. Warning to Humanity: A Second Notice. (2017). *BioScience*, **67**, 1026–1028. https://doi.org/10.1093/biosci/bix125

5. Ripple, W. J., Christopher, W., Newsome, T, M., Phoebe B., and Mommas, W.R. (2020). "World Scientists' Warning of a Climate Emergency." *BioScience* **70**, 8–12. https://doi.org/10.1093/biosci/biz088

6. Planton, S. (2022), in Encyclopedia of the Environment, [online ISSN 2555-0950] url : https://www.encyclopedie-environnement.org/en/climate/average-temperature-earth/

7. Loudon, R. (2000) The Quantum Theory of Light (third ed.). Cambridge University Press. ISBN 0-19-850177-3.

8. Coffey, J. (2015) Temperature of Earth [Online]. http://www.universetoday.com/14516/temperature-of-earth/

9. Bullen, P.S. (2003) Handbook of Means and Their Inequalities, Kluwer Academic Publisher.

10. Riasat, S. (2008) Basics of Olympiad Inequalities, William College, p.24.

11. Feynman R, Leighton R, and Sands M. The Feynman Lectures on Physics. (1964), Addison–Wesley, Boston, Vol. 1.